\documentclass{pasa}
\usepackage{graphicx}
\usepackage{multirow}
\usepackage{lineno}
\usepackage{xcolor}
\usepackage{soul}

\title{Optimising an Array of Cherenkov Telescopes in Australia for the Detection of TeV Gamma-Ray Transients}

\author[Simon Lee et al.]{Simon Lee$^1$, Sabrina Einecke$^1$, Gavin Rowell$^1$, Csaba Balazs$^2$, Jose A. Bellido$^1$, Shi Dai$^3$, Miroslav Filipovi\'c$^3$, Violet M. Harvey$^1$,
Padric McGee$^1$, Peter Marinos$^1$, Nicholas Tothill$^3$, Martin White$^1$
\affil{$^1$School of Physics, Chemistry and Earth Sciences, The University of Adelaide, Adelaide SA 5005, Australia}
\affil{$^2$School of Physics and Astronomy, Monash University, Melbourne VIC 3800, Australia}
\affil{$^3$School of Science, Western Sydney University, Locked Bag 1797, Penrith NSW 2751, Australia}
}%

\jid{PASA}
\doi{10.1017/pas.\the\year.xxx}
\jyear{\the\year}

\usepackage{aas_macros}
\usepackage{hyperref}
\hypersetup{colorlinks,citecolor=blue,linkcolor=blue,urlcolor=blue}

\usepackage{chngcntr}

\begin{document}
\begin{frontmatter}
\maketitle

\begin{abstract}
As TeV gamma-ray astronomy progresses into the era of the Cherenkov Telescope Array (CTA), instantaneously following up on gamma-ray transients is becoming more important than ever.
To this end, a worldwide network of Imaging Atmospheric Cherenkov Telescopes has been proposed.
Australia is ideally suited to provide coverage of part of the Southern Hemisphere sky inaccessible to H.E.S.S. in Namibia and the upcoming CTA-South in Chile.
This study assesses the sources detectable by a small, transient-focused array in Australia based on CTA telescope designs.
The TeV emission of extragalactic sources (including the majority of gamma-ray transients) can suffer significant absorption by the extragalactic background light.
As such, we explored the improvements possible by implementing stereoscopic and topological triggers, as well as lowered image cleaning thresholds, to access lower energies.
We modelled flaring gamma-ray sources based on past measurements from the satellite-based gamma-ray telescope \emph{Fermi}-LAT.
We estimate that an array of four Medium-Sized Telescopes (MSTs) would detect $\sim$24 active galactic nucleus flares >5$\sigma$ per year, up to a redshift of $z\approx1.5$.
Two MSTs achieved $\sim$80--90\% of the detections of four MSTs.
The modelled Galactic transients were detectable within the observation time of one night, 11 of the 21 modelled gamma-ray bursts were detectable, as were $\sim$10\% of unidentified transients.
An array of MST-class telescopes would thus be a valuable complementary telescope array for transient TeV gamma-ray astronomy.
\end{abstract}

\begin{keywords}
Monte Carlo simulations -- IACT -- gamma-ray astronomy -- stereo trigger -- topo trigger -- transients -- AGN -- gamma-ray bursts -- novae -- EBL
\end{keywords}
\end{frontmatter}

\section{INTRODUCTION }

\label{sec:intro}

Extending multi-wavelength studies to GeV and TeV energies has allowed us to probe the nature of the universe’s most extreme sources, environments, and phenomena.
These astronomical domains also act as our highest-energy laboratories for exploring particle physics beyond the standard model.
There are however deficiencies in the field compared to those concerned with other electromagnetic wavelengths.
In particular, there is at present limited capacity to instantaneously follow up on and continuously monitor short-lived, variable, transient phenomena over a 24-hour period.

The flares of active galactic nuclei (AGNs) make up the vast majority of known gamma-ray transients above 10\,GeV \citep{Abdollahi2017}.
The great distance of most AGNs tends to make their TeV emissions subject to significant absorption by the extragalactic background light (EBL).
Gamma-ray bursts (GRBs), now understood to be either from compact stellar mergers or the core collapse of massive stars, are similarly of extragalactic origin and are generally very short lived (lasting milliseconds to, at most, days).
Alongside Galactic transients like novae, pulsars, and binaries, there is a substantial collection of transient events which are either associated with objects of unknown classification, or not associated with any counterpart.
Thoroughly studying these phenomena at high energies is challenging, and there is much opportunity for new discovery.

There are three broad categories of telescopes observing these photon energies, each with their advantages and limitations.
\href{https://glast.sites.stanford.edu/}{\emph{Fermi}-LAT}\footnote{\href{https://glast.sites.stanford.edu/}{glast.sites.stanford.edu}} is a satellite-based direct-detection telescope providing quasi-continuous all-sky monitoring from 20\,MeV to $\sim$1\,TeV \citep{atwood2013pass}.
Its small collection area of $<$1\,m$^2$ \citep{Maldera2021} however results in day-scale time resolution for all but the most extreme sources, and it has low sensitivity to emission above 10\,GeV.
Water Cherenkov detectors, such as \href{https://www.hawc-observatory.org/}{HAWC}\footnote{\href{https://www.hawc-observatory.org/}{hawc-observatory.org}}, \href{https://english.ihep.cas.cn/lhaaso/}{LHAASO}\footnote{\href{https://english.ihep.cas.cn/lhaaso/}{english.ihep.cas.cn/lhaaso}}, and the proposed \href{https://www.swgo.org}{SWGO}\footnote{\href{https://www.swgo.org}{swgo.org}}, measure Cherenkov radiation generated in water from passing charged particles created in gamma-ray- (or cosmic-ray-) induced particle showers in the atmosphere \citep{arteaga2015,zhang2021,abreu2019}.
These benefit from effectively continuous operation, a field of view covering much of the overhead sky, and energy ranges extending up to PeV energies.
They are however not sensitive to short-timescale flux variations, and suffer from poor angular resolution at sub-TeV energies \citep{Wang2018}.
Lastly, Imaging Atmospheric Cherenkov Telescopes (IACTs), such as \href{https://magic.mpp.mpg.de/}{MAGIC}\footnote{\href{https://magic.mpp.mpg.de/}{magic.mpp.mpg.de}}, \href{https://www.mpi-hd.mpg.de/HESS/}{H.E.S.S.}\footnote{\href{https://www.mpi-hd.mpg.de/HESS/}{mpi-hd.mpg.de/HESS}}, and \href{https://veritas.sao.arizona.edu/}{VERITAS}\footnote{\href{https://veritas.sao.arizona.edu/}{veritas.sao.arizona.edu}}, conversely image the Cherenkov radiation from particle showers in air, collected with large (typically segmented) mirror dishes \citep{tridon2010,hess2023,daniel2020}.
These are the most suitable telescopes for observing gamma-ray sources with low flux at short timescales, but have a comparatively small field of view and are restricted to night time observations in favourable weather.

The \href{https://www.cta-observatory.org/}{Cherenkov Telescope Array}\footnote{\href{https://www.cta-observatory.org/}{cta-observatory.org}} (CTA) will be a pair of large IACT arrays in the Northern and Southern Hemispheres.
With class-leading sensitivity and angular resolution and an energy range extending from 20 GeV to $>$300 TeV, it will be the world’s premier gamma-ray observatory \citep{2018CTA}.
Whilst transients are a key science project of the CTA Observatory, its schedule will be heavily influenced by surveys and deep observations.

Currently it is not possible to instantaneously follow up on gamma-ray transients with appropriately sensitive telescopes unless they are conveniently visible to an IACT.
A solution to this deficiency would be a worldwide network of such telescopes, complementing those which already exist.
Indeed, the idea for a large-scale network of IACTs has been proposed multiple times before \citep{Lorenz2005,2009icrc.conf.1452B,Ruhe2019} and the performance of a small IACT array for such observations has similarly been studied \citep{Plyasheshnikov2000,Kifune2001,yoshikoshi2005,Colin2007,Rowell2008,Stamatescu2011}.
This paper is a continuation of previous work on the performance of a hypothetical small IACT array sited in Australia \citep{lee2022}, a crucial and currently un-utilised location for obtaining Southern Hemisphere sky coverage to which it has previously been host \citep{Clay1989, Armstrong1999, Enomoto2002}.

In this study, we have investigated methods of improving the performance of a small IACT array focused on gamma-ray transient detections.
In particular, we have assessed the effect of using alternate telescope triggers to allow summed-pixel-level trigger thresholds to be decreased, ostensibly lowering the array's energy threshold to observe more low-energy gamma rays, without increasing accidental triggers due to light from the night sky background.
Through implementing the best of these results, we explore the array’s suitability for detection and observation of a number of gamma-ray transients as recorded by \emph{Fermi}-LAT in the \emph{Fermi} All-sky Variability Analysis (FAVA) catalogue \citep{Abdollahi2017} and the \emph{Fermi}-LAT Second Gamma-Ray Burst Catalog (FERMILGRB) catalogue \citep{fermi2019}.

\section{METHOD}
\begin{figure}[t]
    \begin{center}
        \includegraphics[width=0.75\columnwidth]{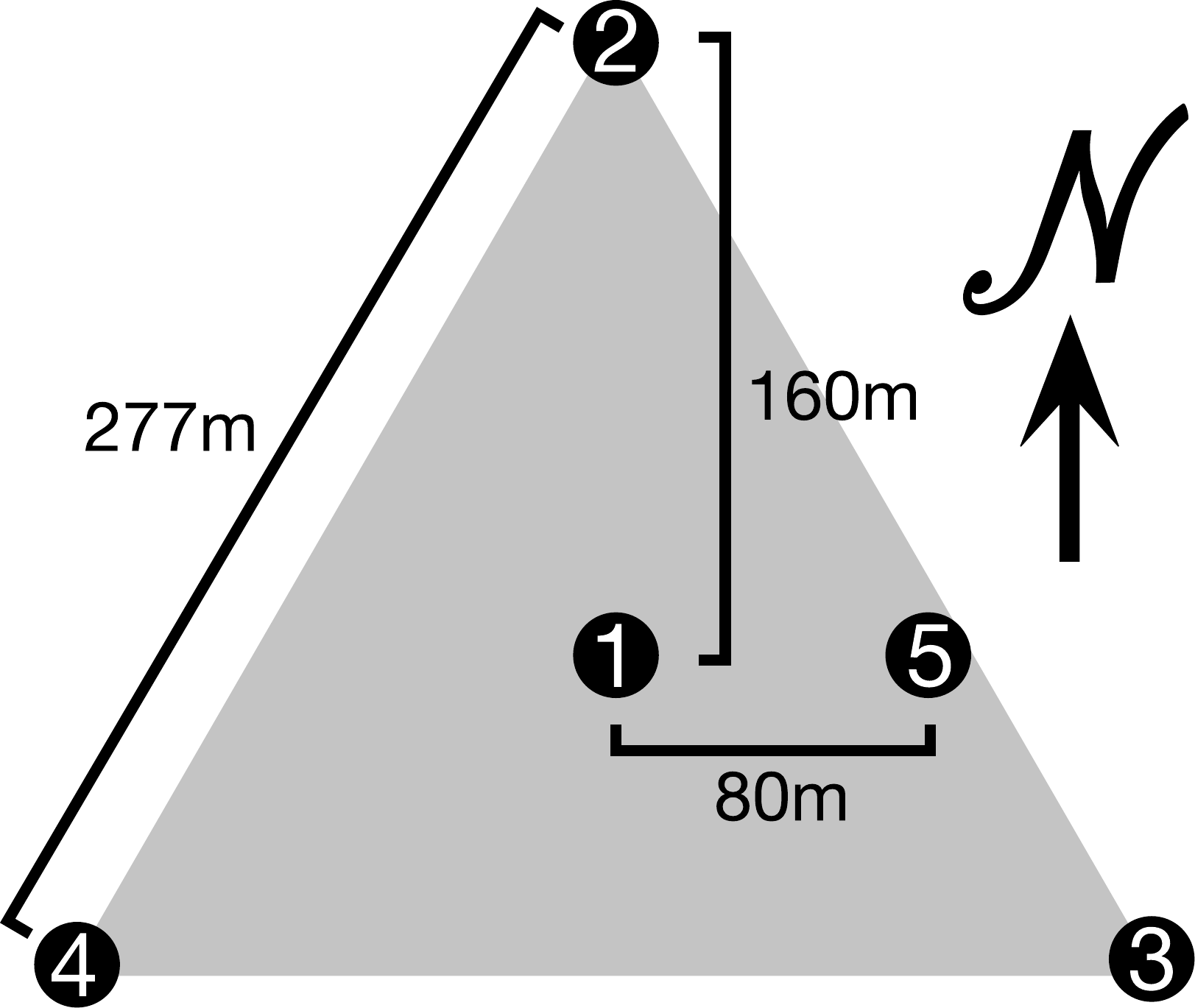}
        \caption{Arrangement of IACTs (shown as numbers) used in simulations, allowing for multiple different configurations of baseline distances and number of telescopes to be studied.}
        \label{tels}
    \end{center}
\end{figure}

This work builds on \cite{lee2022} which describes the scientific motivation, background, and technical methods in more detail.
We simulated telescope arrays as per \autoref{tels}.
Four-telescope arrays used a 277\,m baseline and an inter-telescope distance of 160\,m using telescopes labelled 1, 2, 3, and 4.
We compared two-telescope arrays at baselines of 80\,m (1 \& 5), 160\,m (1 \& 2), and 277\,m (2 \& 3).
This study primarily focused on the detection of transient phenomena at comparatively lower energies (below 1\,TeV), so the simulated energy range of the test data was set to 0.04--3\,TeV, and the spectral index to 2.5 (see \autoref{settings}).

\begin{table*}[t]
\caption{Simulation parameters used in \texttt{CORSIKA}.
Particle energies were drawn from the distribution $dN/dE \propto E^{-\Gamma}$.
Diffuse emission was generated within a ``view cone'' of radius $r_{\mathrm{cone}}$.
Shower core positions were evenly distributed in a circular area of radius $r_{\mathrm{scatter}}$.
Showers were re-used $n_{\mathrm{reuse}}$ times with their core positions varied, providing alternate views of the shower.}
\label{settings}
\centering
\begin{tabular}{lcccccc}
\hline
\begin{tabular}[c]{@{}l@{}}Particle  type\end{tabular} & \begin{tabular}[c]{@{}l@{}}Energies (TeV)\end{tabular} & \begin{tabular}[c]{@{}l@{}}$\Gamma$\end{tabular} & \begin{tabular}[c]{@{}l@{}}$n_{\mathrm{reuse}}$\end{tabular} & \begin{tabular}[c]{@{}l@{}}$r_{\mathrm{cone}}$ (deg)\end{tabular} & \begin{tabular}[c]{@{}l@{}}$r_{\mathrm{scatter}}$ (m)\end{tabular} & \begin{tabular}[c]{@{}l@{}}Total events\end{tabular} \\ \hline
\begin{tabular}[c]{@{}l@{}}Gamma (on-axis, test)\end{tabular} & \begin{tabular}[c]{@{}l@{}}0.04 -- 3\end{tabular} & 2.5 & 20 & 0 & 700 & $11\times10^6$ \\
\begin{tabular}[c]{@{}l@{}}Gamma (diffuse, train)\end{tabular} & \begin{tabular}[c]{@{}l@{}}0.01 -- 300\end{tabular} & 2.5 & 20 & 10 & 1560 & $2.1\times10^9$ \\
Proton (diffuse, train) & \begin{tabular}[c]{@{}l@{}}0.01 -- 300\end{tabular} & 2.5 & 20 & 10 & 1560 & $7.3\times10^9$ \\
Proton (diffuse, test) & \begin{tabular}[c]{@{}l@{}}0.04 -- 3\end{tabular} & 2.5 & 20 & 10 & 800 & $1.1\times10^9$ \\ \hline
\end{tabular}
\end{table*}

We generated Monte Carlo simulations of gamma-ray and proton extensive air showers with \verb|CORSIKA|\footnote{Version 7.7100 with the QGSJET II-04 interaction model.} \citep{1998cmcc} centred at a 20\textdegree\ zenith angle.
The resulting Cherenkov photons were passed into simulations of telescope arrays based on CTA Prod-5\footnote{Version 2020-06-28.} designs \citep{CTAO2021, ctaprod52022} of the 12-metre Medium-Sized Telescope (MST) and 4-metre Small-Sized Telescope (SST), further described in \autoref{cta_appendix}.
We used tools from \verb|ctapipe| \citep{Kosack2021} to extract images from the simulated telescope data, clean them, and perform second-moment Hillas analysis \citep{hillas1985} to create datasets for training and applying event reconstruction models.
Quality cuts were made based on some of the Hillas parameters (described in \autoref{cleaning_section}).
We trained random forest models for reconstructing particle energy and direction and gamma/hadron separation on the diffuse gamma-ray dataset and a portion of the diffuse proton dataset using \verb|aict-tools| \citep{aict-tools}.
These models were applied on the point-source gamma-ray and remaining diffuse proton datasets.
For each of 5 bins per decade in energy, performance cuts of a minimum gamma score (conformity to a gamma ray) and maximum $\theta^2$ (angular separation from reconstructed direction to the expected source position) were chosen for optimal differential sensitivity for a 50-hr point-source observation.

We chose a maximum simulated energy of 3\,TeV, as detection relies on the counted number of gamma rays, and the power law spectra of transient sources typically provide few TeV photons.
We simulated the arrays at sea level, as the availability of suitable high-altitude sites in Australia is limited.
When determining performance cuts, we used the Bayesian-derived method outlined in \cite{Knoetig2014} for significance calculations from on- and off-region counts.
Using this method, performance cuts were only considered if at least 1 on-region and 1 off-region event survived in a given energy bin, instead of 10 as with the method from \citet{Li1983}.

\subsection{Trigger} \label{trigger_section}

To improve the performance of the arrays for gamma-ray transient detection, we sought to decrease the summed-pixel-level trigger threshold (discriminator threshold, DT) to capture the dimmer Cherenkov light of gamma-ray air showers with lower energies.
Typically, this would increase the rate at which the telescope is triggered accidentally by light from the night sky background (NSB), but this effect can be mitigated with alternative trigger configurations.

\begin{figure}[t]
    \begin{center}
        \includegraphics[width=1\columnwidth]{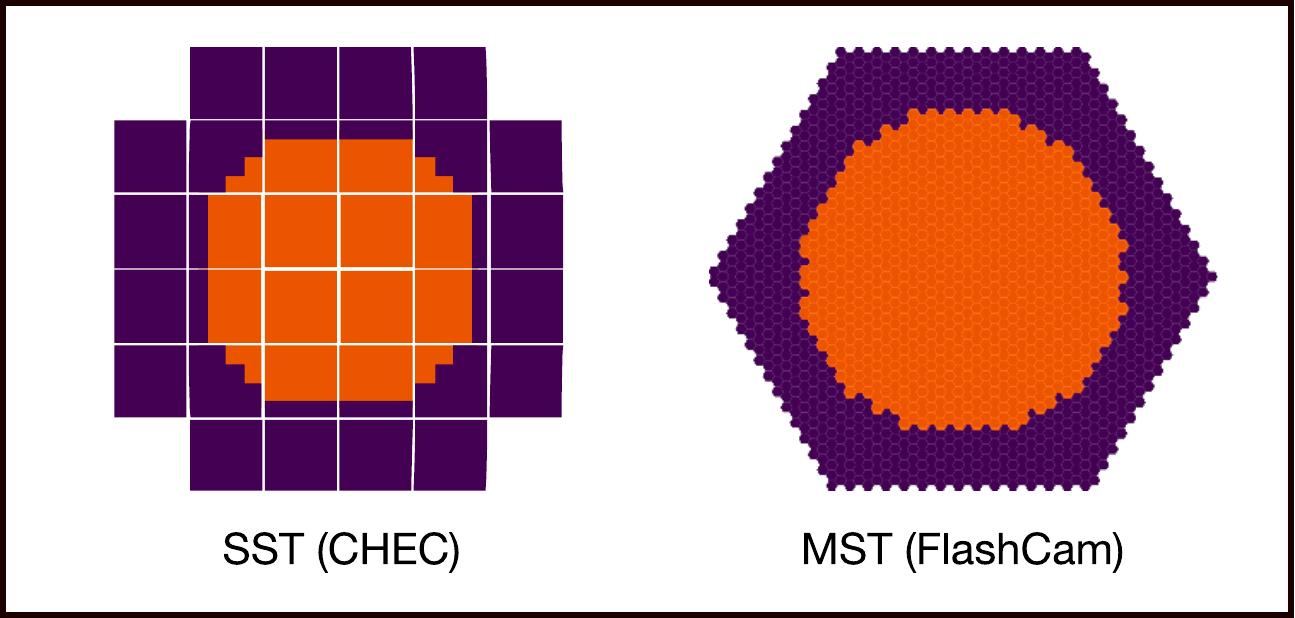}
        \caption{Central circular regions of pixels (orange) used in the SST and MST cameras for the simulated topological trigger.
        See \autoref{cta_appendix} for details on the cameras.}
        \label{topo_trigger_region}
    \end{center}
\end{figure}

In the standard trigger scenario for a given telescope (hereby labelled as ``default'') an event would be saved if a cluster of pixels anywhere in the camera exceeded a given voltage for a given amount of time.
We tested a simple topological (``topo'') trigger, only using a circular region of pixels in the centre of each telescope's camera for potential triggers.
This is a technique first implemented by \citet{lucarelli2003} with the now-decommissioned \href{https://www.mpi-hd.mpg.de/hfm/CT/CT.html}{HEGRA}\footnote{\href{https://www.mpi-hd.mpg.de/hfm/CT/CT.html/}{mpi-hd.mpg.de/hfm/CT}} IACT array.
Low-energy showers are only visible at comparatively small core distances from the array and thus their positions in the camera are clustered closer to the centre.
This technique requires the expected source position to be known, and is not suitable for surveys.
We chose the smallest circular areas that would trigger on all gamma rays seen by the default trigger below 1\,TeV.
These covered 33.6\% of the SST camera and 48.1\% of the MST camera (shown in \autoref{topo_trigger_region}).

In our previous study \citep{lee2022} a simple monoscopic (``mono'') trigger was used, requiring only one telescope in the array to trigger for an event to be saved.
We additionally tested a stereoscopic (``stereo'') trigger, requiring at least two telescopes to be triggered within a short window of time.
This method is currently implemented at telescopes such as MAGIC \citep{magic2021}.
Cherenkov light from an air shower will arrive at two nearby telescopes almost simultaneously, whereas NSB light arrives randomly.
Accidental NSB triggers are thus minimised with smaller trigger window times, however gamma-ray showers may not trigger the array if the time window is too short.
This limit depends on the distance between telescopes and the source position in the sky.
We chose an array trigger window of 50\,ns to minimise NSB triggers while keeping >99\% of simulated gamma rays in all array configurations.

\begin{figure}[t]
    \begin{center}
        \includegraphics[width=0.95\columnwidth]{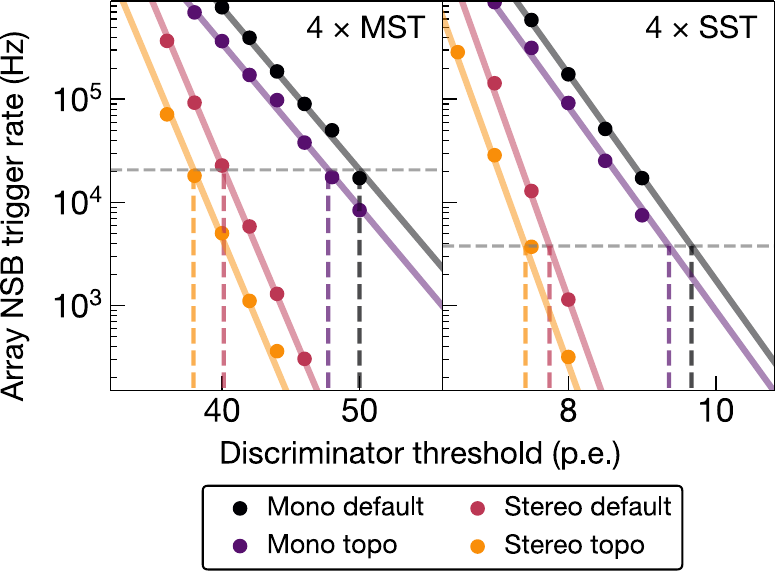}
        \caption{Rate of trigger for the whole array due to night sky background (NSB) for 4$\times$MST and 4$\times$SST arrays as a function of pixel-level discriminator threshold (DT).
        Labels are described in \autoref{trigger_section}.
        The horizontal dashed line indicates the NSB trigger rate using the mono default trigger.
        Vertical dashed lines indicate the corresponding DT to result in the same NSB trigger rate.}
        \label{DT}
    \end{center}
\end{figure}

Lastly, we tested the combination of the stereo and topo triggers, requiring at least two telescopes to trigger in the circular regions of pixel within the 50\,ns trigger window time.
For each combination of telescope size (SST or MST), number of telescopes (2 or 4), and trigger setup, we lowered the DT so as to equalise the rate at which the array triggered on NSB with that of the original (``mono default'') trigger.
The effect of DT on NSB array trigger rate is demonstrated in \autoref{DT}.

\subsection{Cleaning and Quality Cuts} \label{cleaning_section}

We tested multiple lowered cleaning thresholds in the two-stage tail cuts cleaning process.
Pixels survive this process if they exceed a high ``core'' threshold of extracted photoelectrons (p.e.) or a lower ``boundary'' threshold if they are adjacent to a core pixel.
The default core/boundary thresholds used in \citet{lee2022} were 10/5 p.e.\ for MSTs, and 3/1.5 p.e.\ for SSTs.
Reducing these thresholds increases the number of events (of both gamma rays and protons) that survive cleaning, but can introduce noise artefacts.
We also tested excluding the quality cut on total extracted photoelectrons (\texttt{intensity}).
The other quality cuts were kept consistent: \texttt{leakage}\footnote{The ratio (post-cleaning) between summed pixel intensities at the edge of the camera and the total summed intensity.} ($<0.2$), \texttt{surviving pixels} ($\geq6$), and \texttt{number of islands}\footnote{Disjoint clusters of pixels post-cleaning.} ($\leq3$).

\begin{figure*}[t]
    \begin{center}
      \centering
          \includegraphics[width=1\textwidth]{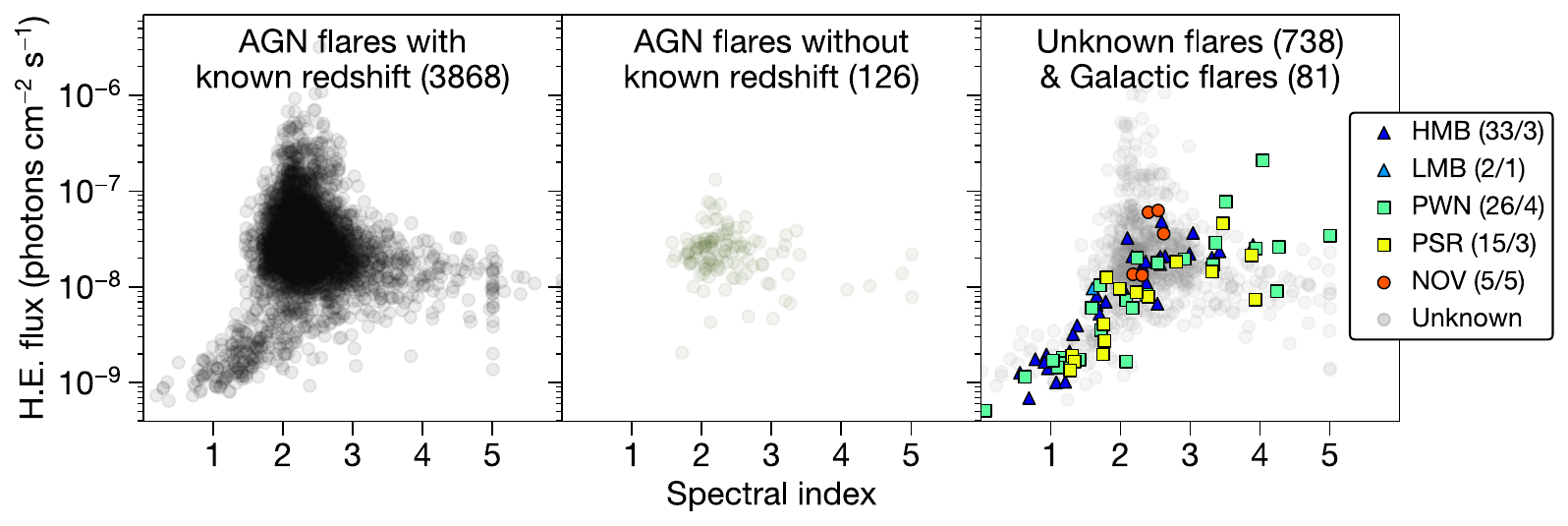}
          \caption{Northern and Southern Hemisphere flares in the FAVA weekly flare catalogue up to \mbox{2023-02-20}, showing their photon flux and spectral index in the \emph{Fermi}-LAT high-energy band (0.8--300\,GeV).
          Only local maxima are included for multi-week flares.
          The Galactic source categories are HMB (high mass binary), LMB (low-mass binary), PWN (pulsar wind nebula), PSR (pulsar), and NOV (nova), listed with their respective number of flares/sources.}
          \label{sources}
    \end{center}
\end{figure*}

\subsection{Transients} \label{transients}

The \emph{Fermi} All-sky Variability Analysis (FAVA) catalogue \citep{Abdollahi2017} is the largest database of GeV gamma-ray transient measurements.
The catalogue currently includes flares from \mbox{2008-08-03} to \mbox{2016-01-04}.
We obtained further flare data up to \mbox{2023-02-20} from the FAVA weekly analysis website\footnote{\href{https://fermi.gsfc.nasa.gov/ssc/data/access/lat/FAVA/}{fermi.gsfc.nasa.gov/ssc/data/access/lat/FAVA/}}.
The catalogue results from an unbiased search of the whole sky using the Large Area Telescope (LAT) on the \emph{Fermi Gamma-ray Space Telescope} satellite.
From the detected photons, a likelihood analysis fit is performed to discover flares with a significance greater than 6$\sigma$ (pre-trial) and to measure their spectral index and flux in two energy bins (100\,MeV--800\,MeV, and 800\,MeV--300\,GeV).

We attempted to associate as many flares as possible to known sources and with published redshifts.
We assigned these through association of the following databases: the table from \citet{foschini2022}, the 4LAC-DR2 catalogue \citep{lac4dr2,lac4}, the 3LAC catalogue \citep{lac3}, BZCAT \citep{bzcat}, and SIMBAD \citep{simbad}.
Aspects of the flares are shown in \autoref{sources}.

When modelling the flares, we assumed that the intrinsic energy spectrum of these sources continued with the same spectral index above 300\,GeV to the simulated maximum of 3\,TeV, and modelled the intrinsic spectral energy distribution of flares as a power law: 
\[\frac{dN}{dE}=N_0 \left(\frac{E}{1\,\mathrm{TeV}}\right)^{-\Gamma}\]
where $N_0$ is the normalisation constant at 1\,TeV and $\Gamma$ is the spectral index.
For flares associated with sources of known redshift, we implemented spectral absorption due to the EBL using the model from \citet{Inoue2013}.
The temporal component of the modelling process is discussed for each source class in subsequent sections.
Once the spectrum of a source was modelled, it was convolved with the effective area of the simulated telescope arrays to estimate cosmic-ray background counts \citep[from simulated protons and from the electron spectrum described in][]{HESS2017} and gamma-ray counts for given observation times.
Finally, we calculated the detection significance using the method described by \citet{Knoetig2014}, and significances greater than 5$\sigma$ were considered a successful detection.

\subsubsection{AGN Flares}

In total there were 9910 FAVA flares considered in this study, of which 7892 had spectral parameters fitted for the high-energy bin (0.8--300\,GeV).
Many sources were detected flaring in consecutive weeks, so flares for a given source were selected only if they had a higher measured flux than in adjacent weeks.
This resulted in 4826 ``unique'' flares, 4088 of which were associated with known sources.
The vast majority (3994, 97.7\%) of these were AGN flares, 3868 of which were from 514 AGNs with published redshifts.
The distribution of spectral parameters of these flares in their respective categories is shown in \autoref{sources}.

To only consider events occurring at relatively small zenith angles (where IACT performance is optimal) for a Southern Hemisphere array at a latitude of 30\textdegree\ South, we restricted the catalogue to sources with a declination between $-60^{\circ}$ and $0^{\circ}$, limiting us to 1694 unique flares from 232 AGN with known redshift.
These flares were modelled with static fluxes over 4-hour and 8-hour observation periods, which are (respectively) conservative and optimistic durations for a single night.
The FAVA catalogue only provides weekly average flux values, and one can assume that a flare would likely rise and decay in brightness over shorter periods of time.
\autoref{flux_cartoon} demonstrates scenarios that result in different relative fluxes 26 hours into a \emph{Fermi}-LAT week, i.e.\ in the middle of a 4-hour observation starting 24 hours into the week.
We modelled AGN flares with static fluxes scaled to 1$\times$, 2$\times$, and 4$\times$ their weekly averages.
A scaling factor of 1$\times$ is a pessimistic model, which either assumes that the flare did not vary in brightness over the course of a week, or was observed some time after the peak of its rapidly varying flux. 
A 2$\times$ scaling factor is conservative, whereas 4$\times$ is optimistic.
For the primary results of this study, we considered an observation time of 4 hours per flare at a flux twice that of the weekly average as a conservative model of AGN observability.

\begin{figure*}[t]
    \begin{center}
      \centering
          \includegraphics[width=1\textwidth]{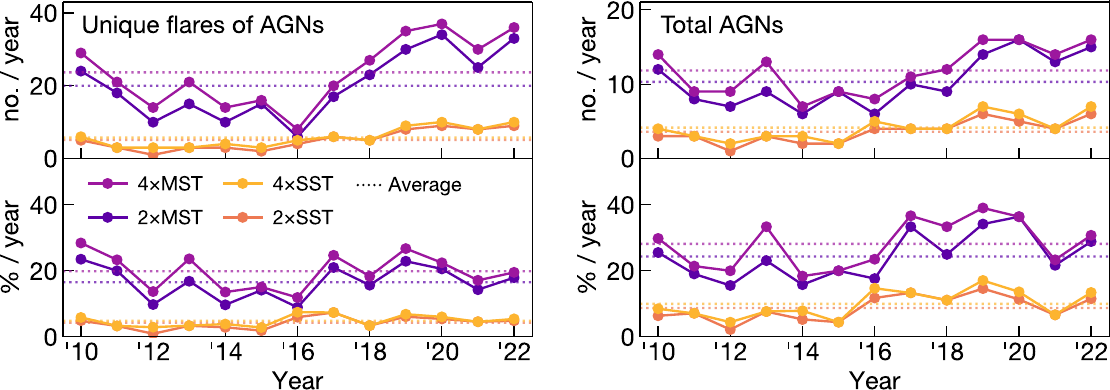}
          \caption{Number (upper) and percentage (lower) of unique flares of AGNs (left) and total AGNs (right) from the FAVA catalogue at $-60^{\circ} \le \mathrm{dec} \le 0^{\circ}$ detectable at $>$5$\sigma$ per year from 2010 to 2022 by the best-performing simulated configurations of two or four MSTs or SSTs (\autoref{best_arrays}).
          Flares were simulated being observed for four hours at twice their weekly average flux.
          }
          \label{thebigone}
    \end{center}
\end{figure*}

\subsubsection{Galactic Flares}\label{galactic_section}

In the selected catalogue, 81 high-energy gamma-ray flares from 16 sources were identified as originating from Galactic sources (shown in \autoref{sources}, right), 12 of which were from 5 sources in the Southern Hemisphere sky.

The gamma-ray binary PSR B1259-63/LS 2883 consists of a pulsar in an eccentric orbit with a bright companion star, orbiting with a period of approximately 3 years and 5 months \citep{HESS2020}.
It was associated with 8 flares in the catalogue from its 2010, 2014, and 2021 periastron passages, however none of these flares surpassed the >4$\sigma$/TS>18 (test statistic) cut in the high-energy band and were only included due to their >6$\sigma$/TS>39 detection at low energies.
Its 2017 periastron is not recorded in the FAVA catalogue, possibly due to an incorrect association with HESS J1303-631.
These flaring episodes are visible to IACTs, so we based our spectral energy distribution model of this binary on the 2014 periastron passage 8-day fit provided in \citet{HESS2020}, with a spectral index of $\Gamma=2.7$ and a static normalisation constant of $N_0 = 2.6\times10^{-12}$\,TeV$^{-1}$\,cm$^{-2}$\,s$^{-1}$.

The classical nova V1324 Scorpii had an outburst in 2012, which was detected by \emph{Fermi}-LAT.
The week of \emph{Fermi}'s detection coincided with a bright plateau in the optical lightcurve \citep{finzell2018}.
We modelled this flare using its parameters in the FAVA catalogue: a spectral index of $\Gamma=2.54$ and a static normalisation constant of $N_0 = 1.645\times10^{-12}$\,TeV$^{-1}$\,cm$^{-2}$\,s$^{-1}$.

V1369 Centauri was a classical nova that erupted in 2013, first discovered by an amateur Australian astronomer \citep{aavso2013} and subsequently followed up by \emph{Fermi}-LAT, which detected rising gamma-ray levels for several successive days \citep{cheung2016}.
We modelled this nova flare using the recorded spectral parameters from \emph{Fermi}-LAT: a spectral index of $\Gamma=2.31$ and a static normalisation constant of $N_0=1.52\times10^{-12}$\,TeV$^{-1}$\,cm$^{-2}$\,s$^{-1}$.

The gamma-ray source HESS J1303-631 was serendipitously discovered in 2004 at TeV energies \citep{HESS2005}, identified as a pulsar wind nebula in 2010 \citep{hess2012}, and later detected by \emph{Fermi}-LAT at GeV energies \citep{acero2013}.
The appearance of HESS J1303-631 as a flaring source in the FAVA catalogue is perhaps erroneous, as it coincides with the (otherwise omitted) 2017 periastron of PSR B1259-63.
This is not improbable given the 1\textdegree\ distance between the two sources and the angular resolution of \emph{Fermi}-LAT.
Nevertheless, \citet{hess2012} provides a model for the spectral energy distribution of the typical state of HESS J1303-631.
We modelled it using their power law fit with no cutoff (the inclusion of which is negligible below 10\,TeV) using a spectral index of $\Gamma=2.44$ and a static normalisation constant of $N_0=5.9\times10^{-12}$\,TeV$^{-1}$\,cm$^{-2}$\,s$^{-1}$.

The \emph{Fermi}-LAT source 4FGL J1813.1-1737e has been associated with the pulsar wind nebula HESS J1813-178.
Its single high-energy flare in the FAVA catalogue exceeded 4$\sigma$ in the high-energy bin, however its inclusion in the database was flagged as potentially being an artefact.
The spectral energy distribution of HESS J1813-178 in a non-flaring state was modelled by \citet{xin2021}, connecting the measured emission from \emph{Fermi}, H.E.S.S., and MAGIC.
We modelled it here based on this fit, with a spectral index of $\Gamma=2.07$ and a static normalisation constant of $N_0=2.55\times10^{-12}$\,TeV$^{-1}$\,cm$^{-2}$\,s$^{-1}$.

\subsubsection{Gamma-Ray Bursts}

The FERMILGRB catalogue \citep{fermi2019} is a list of GRBs detected by \emph{Fermi}-LAT, currently spanning from 2008-08-04 to 2022-06-02.
It provides measured photon fluxes (between 100\,MeV and 100\,GeV) and fitted spectral indices for each GRB, as well as the time window in which the corresponding observation was taken.
The catalogue includes 231 GRBs, of which 120 were in the Southern Hemisphere.
However, only 21 have redshifts provided.
These varied from redshifts of 0.05 to 4.35, with \emph{Fermi}-LAT observation windows lasting between 26 seconds and 3 hours.
We modelled these GRBs using their flux, spectral index (assuming it stays constant from 100\,GeV to 3\,TeV), observation window end time (after which no observable gamma rays are guaranteed), and redshift.
We implemented a decay rate of $t^{-1.2}$, based on the observations of GRB 180720B by \citet{abdalla2019}.
The maximum slew time ($>$30\textdegree\ elevation) for SSTs is 70\,s, and 90\,s for MSTs \citep{ctao2023}, so the observation start times were modelled to begin this long after a trigger from the Gamma-ray Burst Monitor (GBM) on \emph{Fermi}.
A graphical depiction of this modelling process is provided in \autoref{grb_cartoon}.

\section{RESULTS}
The performance impacts of altering telescope array variables are contextual to the desired scientific goals.
Thus, the trigger investigation results are discussed here intertwined with their application for transient discovery.

\subsection{AGN Flares} \label{agn_section}

\begin{figure}[t]
    \begin{center}
        \includegraphics[width=1\columnwidth]{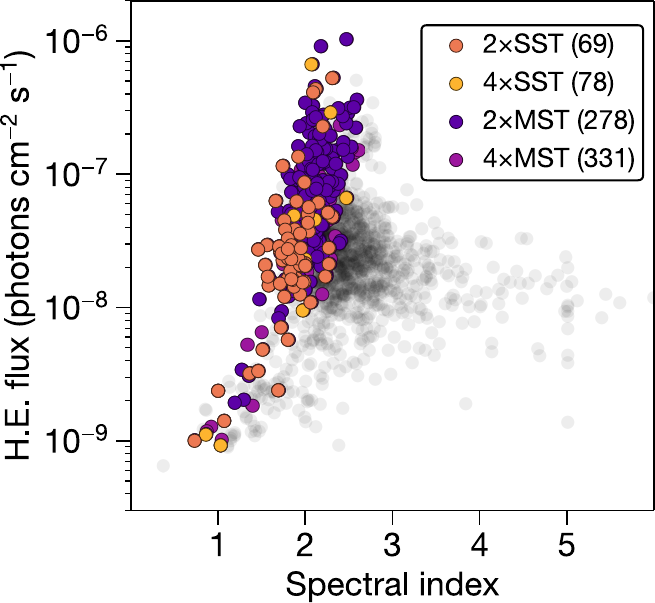}
        \caption{High-energy photon flux and spectral index of unique FAVA AGN flares at $-60^{\circ} \le \mathrm{dec} \le 0^{\circ}$ from August 2008 to February 2023.
        Highlighted points show flares detectable >5$\sigma$ for the configurations used in \autoref{thebigone}, for 4-hour observations and flux 2$\times$ the weekly average, with total flare counts listed in brackets.}
        \label{agns}
    \end{center}
\end{figure}

\autoref{thebigone} shows the yearly number and percentage of detectable flares and sources, with 4 hours of observation per flare and static fluxes 2$\times$ their weekly average, at declinations between $-60^{\circ}$ and $0^{\circ}$.
\autoref{agns} shows their spectral properties.
The redshift distribution of these flares is shown in \autoref{redshifts}.
The telescope configurations and array triggers that resulted in the most detected AGN flares are detailed in \autoref{best_arrays}, and are the arrays used in \autoref{thebigone}.
As presented below, multiple arrays with different configurations performed almost equally well.
\begin{itemize}
  \item For four MSTs, the stereoscopic default trigger performed best with 331 flares detected from 77 sources, trailed only by the stereoscopic topological trigger with 327 flares.
  The farthest detected AGN had a redshift of $z=1.52$.
  \item For two MSTs, the stereoscopic topological trigger with a 160\,m baseline allowed for detection of 278 flares from 69 sources, closely followed by the simplest monoscopic default trigger with 277 flares.
  The farthest detected AGN had a redshift of $z=1.52$.
  \item For four SSTs, the monoscopic topological trigger performed best detecting 78 flares from 33 sources, outperforming the stereoscopic topological trigger which saw 75 flares.
  The farthest detected AGN had a redshift of $z=0.69$.
  \item For two SSTs, the monoscopic default trigger with a 160\,m baseline performed best, detecting 69 of the 1833 flares from 28 unique sources.
  Using a monoscopic topological trigger, or a monoscopic default trigger with an 80\,m baseline, resulted in 68 and 67 flares, respectively.
  The farthest detected AGN had a redshift of $z=0.54$.
\end{itemize}

\begin{table}[t]
\caption{Configurations of the best-performing arrays of two and four SSTs and MSTs for observing AGN flares in the Southern Hemisphere.
A limited number of cleaning thresholds (listed as core/boundary thresholds) were tested.
For comparison, thresholds used in \citet{lee2022} were 10/5 p.e.
for MSTs and 3/1.5 p.e. for SSTs.}
\label{best_arrays}
\centering
\begin{tabular}{r|cccc|}
\cline{2-5}
\multicolumn{1}{l|}{}& \multicolumn{1}{c|}{\begin{tabular}[c]{@{}c@{}}Baseline\\ dist. (m)\end{tabular}} & \multicolumn{1}{c|}{\begin{tabular}[c]{@{}c@{}}Stereo\\ trig.\end{tabular}} & \multicolumn{1}{c|}{\begin{tabular}[c]{@{}c@{}}Topo\\ trig.\end{tabular}} & \begin{tabular}[c]{@{}c@{}}Clean.\\ thresh. (p.e.)\end{tabular} \\ \hline
\multicolumn{1}{|r|}{\begin{tabular}[c]{@{}r@{}}4×\\ MST\end{tabular}} & 277& Yes & No& 5.4/2.7\\ \hline
\multicolumn{1}{|r|}{\begin{tabular}[c]{@{}r@{}}2×\\ MST\end{tabular}} & 160& Yes& Yes& 5.5/2.8\\ \hline
\multicolumn{1}{|r|}{\begin{tabular}[c]{@{}r@{}}4×\\ SST\end{tabular}} & 277& No& Yes & 1.20/0.60\\ \hline
\multicolumn{1}{|r|}{\begin{tabular}[c]{@{}r@{}}2×\\ SST\end{tabular}} & 160 & No& No& 0.72/0.36\\ \hline
\end{tabular}%
\end{table}

Within the assumptions of this modelling, there was no significant benefit to using either a stereoscopic or topological trigger for an array with only two telescopes.
A 4$\times$SST array using a topological trigger detected $\sim$9\% more AGN flares than when using the default trigger setup, but a stereoscopic trigger provided no increase.
A 4$\times$MST array with a stereoscopic trigger detected $\sim$4\% more AGN flares than with the default trigger, but a topological trigger was not beneficial.
Compared to the arrays simulated in \citet{lee2022}, the most significant tested factor that improved the detection of transients was the reduction of thresholds used in the two-stage cleaning process by between 45\% and 76\%.
This accounted for an increase in observable AGN flares of $\sim$30-40\%.

\autoref{n_flares} presents the average and maximum number of yearly flares detectable by these best-performing arrays under different assumptions of observation time and flux scaling.
Under the assumptions used in \autoref{thebigone} of  4-hour observation per flare at fluxes $2\times$ their weekly average, an array of four MSTs would detect an average of $\sim$25 of the AGN flares detected per year by \emph{Fermi}-LAT.
This accounts for $\sim$19\% of the high-energy flares at declinations between $-60^{\circ}$ and $0^{\circ}$ detected by \emph{Fermi}-LAT per year, as well as an average of $\sim$13 ($\sim$27\%) unique sources per year.
The maximum number of flares detected in a calendar year was 40 (in 2020).
A more conservative estimate of the observed flux ($1\times$ weekly average) results in $\sim$15 flares detected per year with 4 hours of observation per transient, or $\sim$20 per year with 8 hours per transient.
Across the different models, an array of two MSTs detected $\sim$80-90\% of the flares detected by four MSTs.
An array of SSTs, with their substantially higher energy threshold, detected approximately one quarter to one third as many flares as an equivalent MST array.

\begin{table}
\caption{Number of FAVA AGN flares at $-60^{\circ} \le \mathrm{dec} \le 0^{\circ}$ detectable per year from 2010 to 2022, depending on flare observation time, assumed flux of flares relative to weekly average (×1, ×2, ×4), and array setup.}
\label{n_flares}
\resizebox{\columnwidth}{!}{%
\centering
\begin{tabular}{ccccc|ccc}
 &  & \multicolumn{3}{c|}{\textbf{4 hour obs.}} & \multicolumn{3}{c}{\textbf{8 hour obs.}} \\ \cline{3-8} 
 & \multicolumn{1}{c|}{} & \multicolumn{1}{c|}{×1} & \multicolumn{1}{c|}{×2} & ×4 & \multicolumn{1}{c|}{×1} & \multicolumn{1}{c|}{×2} & \multicolumn{1}{c|}{×4} \\ \hline
\multirow{2}{*}{\begin{tabular}[c]{@{}c@{}}4×\\ MST\end{tabular}} & \multicolumn{1}{c|}{Avg.} & 14.5 & 23.7 & 36.8 & 19.2 & 29.9 & \multicolumn{1}{c|}{45.1} \\
 & \multicolumn{1}{c|}{Max} & 27 & 37 & 59 & 34 & 51 & \multicolumn{1}{c|}{72} \\ \hline
\multirow{2}{*}{\begin{tabular}[c]{@{}c@{}}2×\\ MST\end{tabular}} & \multicolumn{1}{c|}{Avg.} & 11.5 & 20.0 & 31.1 & 16.0 & 25.0 & \multicolumn{1}{c|}{38.7} \\
 & \multicolumn{1}{c|}{Max} & 23 & 34 & 53 & 32 & 40 & \multicolumn{1}{c|}{60} \\ \hline
\multirow{2}{*}{\begin{tabular}[c]{@{}c@{}}4×\\ SST\end{tabular}} & \multicolumn{1}{c|}{Avg.} & 3.6 & 5.8 & 8.3 & 4.7 & 6.8 & \multicolumn{1}{c|}{10.5} \\
 & \multicolumn{1}{c|}{Max} & 7 & 10 & 17 & 9 & 14 & \multicolumn{1}{c|}{20} \\ \hline
\multirow{2}{*}{\begin{tabular}[c]{@{}c@{}}2×\\ SST\end{tabular}} & \multicolumn{1}{c|}{Avg.} & 3.2 & 5.1 & 7.1 & 4.2 & 6.0 & \multicolumn{1}{c|}{8.8} \\
 & \multicolumn{1}{c|}{Max} & 7 & 9 & 14 & 9 & 11 & \multicolumn{1}{c|}{19} \\ \cline{3-8} 
\end{tabular}%
}
\end{table}

\subsection{Unknown Flares} \label{unknown_section}

\begin{figure}[t]
    \begin{center}
        \includegraphics[width=1\columnwidth]{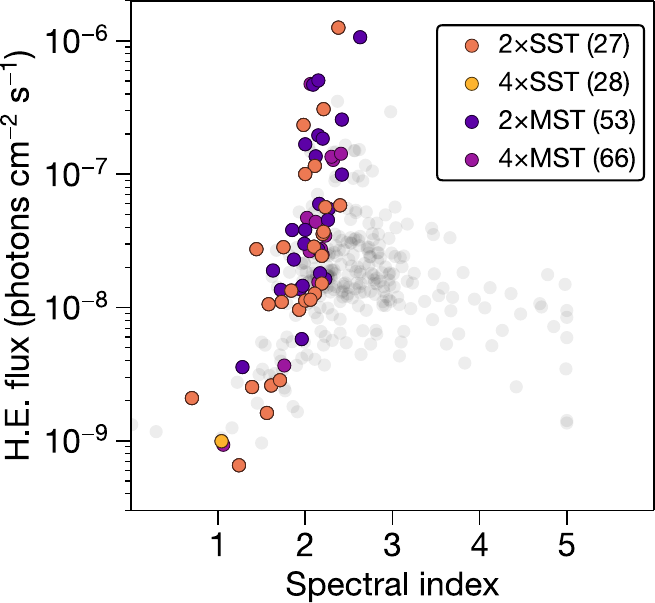}
        \caption{High-energy photon flux and spectral index of FAVA flares at $-60^{\circ} \le \mathrm{dec} \le 0^{\circ}$ with unknown classifications from August 2008 to February 2023.
        Highlighted points show flares detectable >5$\sigma$ for the configurations used in \autoref{thebigone}, for 4-hour observations and flux 2$\times$ the weekly average, with total flare counts listed in brackets.
        }
        \label{unknowns}
    \end{center}
\end{figure}

A large number of flares in the FAVA catalogue are either associated with sources whose nature is unknown, or not associated with any sources at all.
These account for 738 ($\sim$15\%) of the unique flares chosen for this study.
To provide an estimate of their detectability, we modelled these flares in the same way as the AGN flares in \autoref{agn_section}.
We randomly assigned redshifts according to the redshift distribution of AGN flares.
\autoref{unknowns} shows the number and distributions of detectable flares of unknown origin at declinations between $-60^{\circ}$ and $0^{\circ}$.
Unsurprisingly, the distribution of flares whose source category is unknown skews towards a lower photon flux.
As such, less than 10\% of these flares were detectable to the simulated array of four MSTs, compared to the $\sim$19\% for AGN flares.
This would account for roughly another 5 flares per year for four MSTs, 4 for two MSTs, and 2 for the SST arrays.

\subsection{Galactic Flares} \label{galacticflares}

The best trigger setups for observing Galactic flares were as follows:

\begin{itemize}
  \item 4$\times$MST: Mono default trigger.
  \item 2$\times$MST: Stereo default trigger, 277\,m baseline, followed by mono default.
  \item 4$\times$SST: Mono topo trigger, closely followed by stereo topo.
  \item 2$\times$SST: Mono default trigger, 80\,m baseline.
\end{itemize}

In this context of observing Galactic high-energy transients, using a topological trigger noticeably improved the performance of a 4$\times$SST array.
A stereoscopic trigger showed an improvement for a 2$\times$MST array.
For arrays of two SSTs or four MSTs, the performance of the default monoscopic trigger (with lowered cleaning thresholds) was not surpassed.
These results are consistent with the best-performing arrays for detecting AGN flares, except for the 4$\times$MST array wherein a monoscopic trigger benefited from the lack of EBL absorption at high-energies to outperform a stereoscopic trigger.

\autoref{non_agn_time} lists the observation times required for a 5$\sigma$ detection of the Galactic flares whose models are described in \autoref{galactic_section}, using the previously listed array configurations.
An array of two or four MSTs would be very suitable for the observation of the regular flares from PSR B1259-63, whereas equivalent SST arrays require a four-fold increase in observation time.
Both of the flares from the novae V1324 Sco and V1369 Cen would be detectable in one night by MST arrays.
Given the known day-scale longevity and variability of these flares, an SST array would be sufficient to detect them over multiple nights (assuming a static flux equal to their weekly average), but is perhaps unfavourable for time-dependent studies.
Assuming that HESS J1303-631 is indeed a flaring source, its variability would be exceedingly detectable by all simulated arrays.
At its typical flux, HESS J1813-178 would be detectable in one night by MST arrays but not SSTs.
If the source flux is variable at high energies, a flux 40\% above the weekly average would be enough for the SST arrays to achieve a 5$\sigma$ detection within 4 hours.

\begin{table}
\caption{Observation time required for 5$\sigma$ detections by different array configurations of Southern Hemisphere high-energy transient Galactic sources in the FAVA catalogue.
Sources were modelled with static fluxes as per \autoref{galactic_section}.}
\label{non_agn_time}
\centering
\begin{tabular}{rc|c|c|c}
\multicolumn{1}{l}{} & \begin{tabular}[c]{@{}c@{}}4×\\ MST\end{tabular} & \begin{tabular}[c]{@{}c@{}}2×\\ MST\end{tabular} & \begin{tabular}[c]{@{}c@{}}4×\\ SST\end{tabular} & \begin{tabular}[c]{@{}c@{}}2×\\ SST\end{tabular} \\ \cline{2-5} 
\multicolumn{1}{r|}{\begin{tabular}[c]{@{}r@{}}PSR\\ B1259-63\end{tabular}} & 1.5\,hr & 1.9\,hr & 7.0\,hr & \multicolumn{1}{c|}{8.5\,hr} \\ \hline
\multicolumn{1}{r|}{\begin{tabular}[c]{@{}r@{}}V1324\\ Scorpii\end{tabular}} & 5.5\,hr & 6.0\,hr & 17.0\,hr & \multicolumn{1}{c|}{18.5\,hr} \\ \hline
\multicolumn{1}{r|}{\begin{tabular}[c]{@{}r@{}}V1369\\ Centauri\end{tabular}} & 9.5\,hr & 8.0\,hr & 19.0\,hr & \multicolumn{1}{c|}{20.0\,hr} \\ \hline
\multicolumn{1}{r|}{\begin{tabular}[c]{@{}r@{}}HESS\\ J1303-631\end{tabular}} & 35\,min & 35\,min & 1.6\,hr & \multicolumn{1}{c|}{1.8\,hr} \\ \hline
\multicolumn{1}{r|}{\begin{tabular}[c]{@{}r@{}}HESS\\ J1813-178\end{tabular}} & 5.0\,hr & 3.5\,hr & 7.0\,hr & \multicolumn{1}{c|}{7.0\,hr} \\ \hline
\end{tabular}
\end{table}

The anomaly of the 4$\times$MST array performing worse than the 2$\times$MST array in the cases of V1369 Cen and HESS J1813-178 reveals a deficiency in our method for optimising performance cuts in this use case.
The gamma score and $\theta^2$ cuts were optimised for sensitivity independently in each energy bin for 50-hour observations, rather than for overall speed of transient detection.

\subsection{Gamma-Ray Bursts}

For detecting the 21 Southern Hemisphere GRBs with known redshift detected by \emph{Fermi}-LAT, the performances of the simulated arrays were as follows:

\begin{itemize}
  \item 4$\times$MST: 11 GRBs (52\%) detected >5$\sigma$. Best significance with a stereo trigger. No fewer GRBs detected with monoscopic.
  \item 2$\times$MST: 11 GRBs (52\%) detected >5$\sigma$. Best significance with stereo-topo trigger. No fewer with mono.
  \item 4$\times$SST: 5 GRBs (24\%) detected >5$\sigma$. Best with stereo trigger. Similar with mono.
  \item 2$\times$SST: 4 GRBs (19\%) detected >5$\sigma$ with all configurations.
\end{itemize}

\begin{figure}[t]
    \begin{center}
        \includegraphics[width=1\columnwidth]{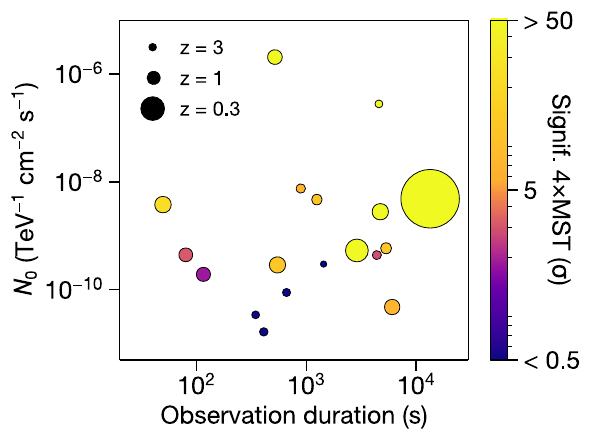}
        \caption{Detectability (with 4$\times$MST) and characteristics of Southern Hemisphere GRBs with known redshift based on \emph{Fermi}-LAT data from August 2008 to June 2022.
        Observation durations were from 90\,s post-trigger from \emph{Fermi}-GBM until the end of the corresponding \emph{Fermi}-LAT observation.
        Depicted are the normalisation constant $N_0$ at 1\,TeV, redshift (inverse to point size), and detection significance from a 4$\times$MST array (colour).
        }
        \label{grbs}
    \end{center}
\end{figure}

These results are consistent with the AGN observability results, which are also strongly dependent on redshift.
The monoscopic default trigger was sufficiently performant for all arrays, but a stereo trigger provided improvements to the 4$\times$SST and 4$\times$MST arrays.
The MSTs performed best with a stereo-topo trigger.
\autoref{grbs} depicts the observability of these Southern Hemisphere GRBs for a 4$\times$MST array with a stereo trigger.
The highest redshift GRB observable by a 4$\times$MST array was GRB181020A, with a redshift of 2.938.
This was first observed by \emph{Fermi}-LAT 75 minutes after the initial trigger from \emph{Swift}.
Whilst this long time delay may cause our retroactive decay estimation to be exaggerated, the X-ray lightcurve from \emph{Swift}-XRT does show a strong signal and broadly consistent decay rate from this GRB \citep{swiftgrb}.
We estimated that GRB150403 and GRB100728 would be observable at their respective redshifts of 2.06 and 1.567, whose \emph{Fermi}-LAT observations started 7 and 4 minutes respectively post-trigger.
Regardless, a large number of the GRBs detectable by \emph{Fermi}-LAT could additionally be detected by a small IACT array consisting of MSTs.

\section{DISCUSSION}
The most significant studied factor to influence a given IACT array's performance for transient discovery was the image cleaning thresholds.
Allowing more pixels to survive, and thus more low-energy events, increased AGN flare detection by $\sim$30-40\% despite the increase in cosmic-ray event survival and the negative effects on event reconstruction.
Additionally, stereoscopic and topological triggers (whether individually or in tandem) were shown to provide measurable but not radical improvements to performance in some contexts.

With respect to AGN flare follow-up, it is clear that an IACT array in Australia would be a productive complement to existing telescopes.
Even with an estimate of 4-hour observations of flares at their weekly average flux, an array of two MSTs would successfully observe $\sim$12 AGN flares per year on average, with a maximum of 23 detections (shown in \autoref{n_flares}).
With the more optimistic estimates of a 4$\times$MST array averaging 45 flare detections a year with a maximum of 72 flares, it is likely that the output of a transient-focused IACT array would fall somewhere in between.
\citet{mura2020} performed a related study, assessing the gamma-ray transient detection capabilities of CTA-South.
Focusing on flares with substantial flux above 10\,GeV as measured by \emph{Fermi}-LAT, they remarked that ``\emph{CTA will be able to detect dozens of transients per year with just 1\,hr of observation}'' given appropriately rapid positional triggers.
Given the differences in event selection criteria and the sensitivity of CTA-South, our results are consistent.

As highlighted in \autoref{agn_section}, it is worthwhile noting that a 2$\times$MST array detected $\sim$80-90\% of the AGN flares detected by an array of four MSTs.
Whilst the additional flare detections, increased angular resolution, and general performance improvements from extra telescopes is obvious, this is a significant consideration if budgetary or space constraints are a concern.
Using SSTs for this kind of transient-focused array is possible, but undesirable given the roughly four-fold decrease in observable AGN flares compared to MSTs.

In addition to the previously discussed gamma-ray transient phenomena, tidal disruption events (TDEs), wherein a star is ripped apart by a massive black hole, have been receiving increased attention as their detection has become more prevalent.
Some have been seen to produce relativistic jets with non-thermal radio and X-ray emission \citep{Andreoni2022}, and \citet{stein2021} has associated one with the production of PeV neutrinos.
\emph{Fermi}-LAT has as yet been unsuccessful in detecting gamma rays from TDEs, however under favourable conditions, an IACT is expected to observe gamma rays if they were produced by hadronic interactions \citep{suvi2021}.
Successfully following up on triggers from their non-thermal radio emission with a worldwide IACT network would help distinguish between particle acceleration methods of TDEs and model their contribution to the diffuse cosmic neutrino flux.

Regarding operational feasibility, the maximum number of detectable AGN flares per year by a 4$\times$MST array was 37 in \autoref{thebigone}.
With an additional $\sim$5 flares per year from unknown sources (as per \autoref{unknown_section}), 168 hours of observation time per year would be required to observe this many flares.
Given a conservative estimate of at least 1000 hours of uptime per year for an IACT array, this is easily achievable.
Additional time would then be available for longer follow-up observations, continuous source monitoring alongside H.E.S.S.\ and CTA-South, responding to weaker triggers, and performing independent searches for gamma-ray transients.

\subsection{Selection Criteria} \label{selection}

A noteworthy distinction between our study and \citet{Abdollahi2017} lies in the flare selection criteria.
We included high-energy flares in our analysis even if their detection significance at high energies was not substantial.
Conversely, \citet{Abdollahi2017} omitted flares from their high-energy plots and analysis that did not pass the 4$\sigma$ or TS>18 cuts in the high-energy bin.
We justify the inclusion of these flares by assuming that if they passed the 6$\sigma$/TS>39 cut in low energies, their high-energy spectral measurements were sufficiently significant, even if they were not independently significant.
When implementing the high-energy significance requirement on the AGN flare database, we observed a negligible reduction in detected AGN flares (shown in \autoref{thebigone_b}).
In the case of a 4$\times$MST array, the average detection rate of AGN flares reduces from 23.7 to 22.3 per year.
As visible in \autoref{sources_b} (upper), the majority of these now-excluded flares are in the hard-spectrum/low-flux regime (bottom left corner).
These flares would be easily detectable in the low-energy band but provide low statistics at high energies due to the power-law nature of the spectra.
For Galactic flares, the total population decreased from 81 to 15 with this cut (shown in \autoref{sources_b}, lower).
Of the Southern Hemisphere Galactic flares, this would exclude PSR B1259-63, however the relevance of this source is discussed in \autoref{galacticflares}.

\subsection{Limitations}
\definecolor{lightblue}{rgb}{.69,.94,0.94}

It is important to recognise the proof-of-concept nature of this study, and the limitations of the methods presented.
The following aspects potentially contribute to overestimates of detection rate.
The assumption of flares being observed at fluxes twice their weekly average in \hbox{\autoref{thebigone}} cannot of course be guaranteed for all flares, hence the inclusion of \autoref{n_flares} for reference.
Additionally, the duration of some flares may be too short, or they may occur at a time of year, so as to not be visible from a given site during clear, dark nighttime (which occurs $\sim$15\% of the year).
The lightcurve profiles of AGN flares vary substantially, however more time-resolved gamma-ray lightcurve data is limited and not representative of the thousands of included flares.
Sources were simulated at a zenith angle of 20\textdegree, and whilst AGN flares were selected with declinations between \hbox{0\textdegree\ } and \hbox{$-60$\textdegree\ }(allowing for low-zenith observations), the dependence of zenith angle was not investigated.
Larger zenith angles result in higher energy thresholds, as does increased moon brightness (and thus NSB) which was not accounted for.
We modelled flares as though their spectral indices were constant past the 300\,GeV energy maximum of \emph{Fermi}-LAT up to the maximum simulated energy of 3\,TeV, however it is entirely possible that some had an intrinsic spectral cutoff that was neither measurable nor modelled.
Our study estimates 11 GRBs to be detectable over 14 years by an MST array, given immediate follow-up.
Our model implicitly assumes that the GRBs would be observed in a clear, dark, night sky at a low zenith angle.
Considering a typical IACT duty cycle of 15\%, our detection rate is comparable with those from \hbox{H.E.S.S.} and MAGIC\footnote{H.E.S.S.\ has detected 2, MAGIC has detected 4 \citep{berti2023}}.

The following aspects potentially contribute to underestimates of detection rate.
Many flares exhibit day-, hour-, or minute-scale variation.
As discussed, it is thus reasonable to assume that some flares could be observed at flux levels several times higher than their weekly averages, given that the weekly flux would be compressed to a shorter time period.
The FAVA catalogue is by nature a conservative measure of the brightness and prevalence of gamma-ray transients, and some flares visible to IACTs may not be included at all, such as orphan TeV flares (which do not appear at other wavelengths).
Furthermore, of flares that vary more slowly, many remain visible for consecutive nights, providing the opportunity for multiple observations and a longer total observation time than the 4 hours modelled in \autoref{thebigone} (assuming there are significantly fewer flares than operational nights).
Additionally, other telescopes operating at different wavelengths may detect transients that \emph{Fermi}-LAT misses due to weak fluxes at GeV energies, providing even more sources for follow-up than in the FAVA catalogue.
In this study we did not investigate the use and effects of convergent pointing, wherein telescopes are pointed towards a point in the atmosphere rather than directly at the source to maximise the stereo overlap for low-energy events \citep[such as in][]{lucarelli2003}, or divergent pointing, which increases the effective field-of-view of the array.
Lastly, the optimisation of array configuration, trigger mechanisms, cleaning thresholds, and quality cuts in this study was not exhaustive.

\section{CONCLUSION}
This study aimed to investigate the suitability of a small Imaging Atmospheric Cherenkov Telescope (IACT) array sited in Australia for dedicated follow-up and detection of gamma-ray transients.
Additionally, we explored the improvements in performance achievable by implementing alternative array trigger methods and event reconstruction parameters.

Arrays of Small-Sized Telescopes (SSTs) and Medium-Sized Telescopes (MSTs) based on designs from the Cherenkov Telescope Array (CTA) were simulated with a variety of configurations.
Flares from active galactic nuclei (AGNs) at declinations between $-60^{\circ}$ and $0^{\circ}$ with known redshift were then modelled based on data from the \emph{Fermi} All-sky Variability Analysis (FAVA) catalogue.
Assuming 4-hour observations with flux twice the measured weekly average, we estimate that an array of four MST-class telescopes would detect an average of $\sim$24 AGN flares above 5$\sigma$ from $\sim$12 distinct sources per year,  with a maximum of 37 flares.
A 2$\times$MST array under the same conditions would detect an average of $\sim$20 flares and a maximum of 34, up to 90\% that of a 4$\times$MST array.
Equivalent arrays with SST-class telescopes were estimated to observe only one quarter the flares of their MST counterparts, which is consistent with their higher energy threshold.

Compared to the arrays we simulated in \citet{lee2022}, we found that the largest improvement in transient detection performance was from reducing the thresholds in the two-stage cleaning process by anywhere from 45\% to 76\%.
This accounted for an increase in observable AGN flares of $\sim$30--40\%.
Implementing a stereoscopic trigger, wherein at least two telescopes in an array must trigger within a 50\,ns window, allowed for adjustments resulting in a 4$\times$MST array detecting $\sim$4\% more AGN flares.
A topological trigger that restricts trigger pixels to a central circular area in the camera allowed a 4$\times$SST array to detect $\sim$9\% more AGN flares.
Neither of these methods resulted in improvements to the simulated two-telescope systems.

For the few known Southern Hemisphere Galactic gamma-ray transients, such as PSR B1259-63 and the novae V1324 Scorpii and V1369 Centauri, 5$\sigma$ detections were achievable within one night's observation.
Whilst requiring fortuitous circumstances to successfully follow-up, over half of the 21 modelled gamma-ray bursts were detectable by MST-class telescope arrays (assuming appropriate visibility and a 90-second trigger response time).
The flares of sources as yet unidentified in the FAVA catalogue would be prime targets for this kind of array, and our estimates suggest $\sim$4-5 of these flares per year could be successfully observed with an array of MST-class telescopes.

Establishing a Southern Hemisphere, transient-focused IACT array in Australia would complement CTA to facilitate further, deeper studies into the most extreme phenomena in our universe.

\begin{acknowledgements}
This work was conducted in the context of the CTA Consortium.
We gratefully acknowledge the work of the CTA simulation and telescope teams for the MST and SST configurations.
This work was supported with supercomputing resources provided by the Phoenix HPC service at the University of Adelaide.
S.L. gratefully acknowledges support through the provision of an Australian Government Research Training Program Scholarship.
This paper has gone through internal review by the CTA Consortium.
\end{acknowledgements}

\bibliographystyle{pasa-mnras}
\bibliography{bibl}

\begin{appendix}
\counterwithin{figure}{section}
\counterwithin{table}{section}

\newpage

\section{Flare decay examples}

\begin{figure}[!ht]
    \begin{center}
          \centering
          \includegraphics[width=1\columnwidth]{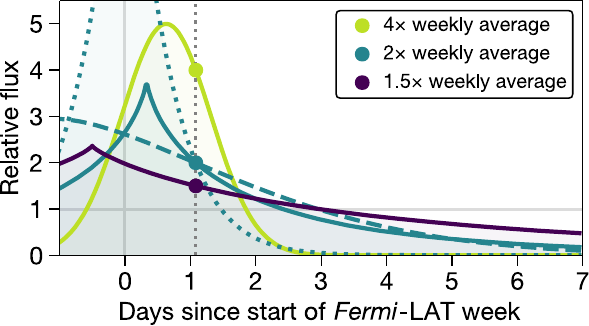}
          \caption{Demonstration of flare shapes that result in particular relative fluxes 26 hours in to a \emph{Fermi}-LAT week (i.e. in the middle of a 4-hour observation starting after 24 hours)
          All flare shapes have the same total flux integrated over the whole week.
          Dashed and dotted lines demonstrate alternate shapes for the same $2\times$ relative flux.}
          \label{flux_cartoon}
    \end{center}
\end{figure}

Shapes in \autoref{flux_cartoon} are based on a symmetric ``generalised Gaussian'' model from \citet{norris1996}, with the form:
\[F(t) = Ae^{-\left(\mathrm{ln}(2)^{\left(1/K\right)}|t-t_0| / \tau_0 \right)^K}\]
where $x_0$ is the peak time, $\tau_0$ is the time from peak to half maximum, and $K$ controls the peak sharpness.

\section{AGN redshift}

\begin{figure}[h!]
    \begin{center}
        \includegraphics[width=1\columnwidth]{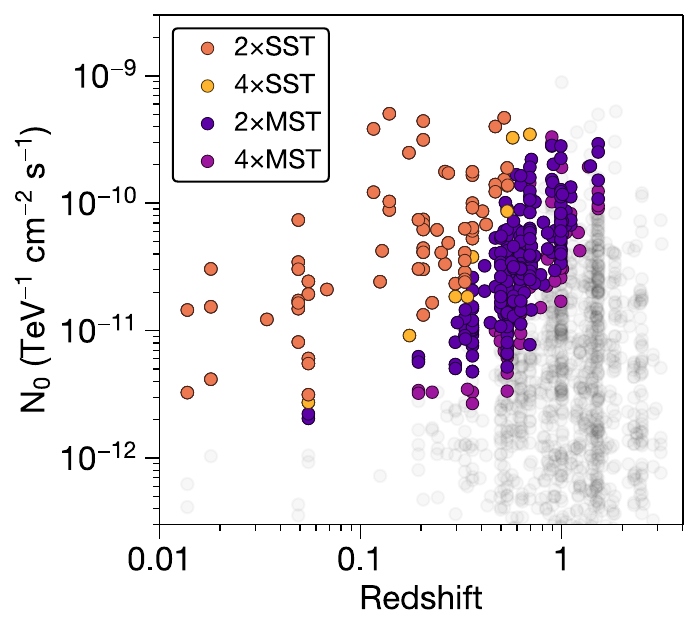}
        \caption{Normalisation constant ($N_0$) at 1\,TeV and redshift of AGN flares from \autoref{agns}, highlighting those detectable at >$5\sigma$.}
        \label{redshifts}
    \end{center}
\end{figure}

\newpage

\section{High-energy significance cut}

The following plots require all flares to pass at least $4\sigma$ or TS>18 in the high-energy (0.8-300\,GeV) FAVA bin.

\begin{figure}[h!]
    \begin{center}
          \centering
          \includegraphics[width=1\columnwidth]{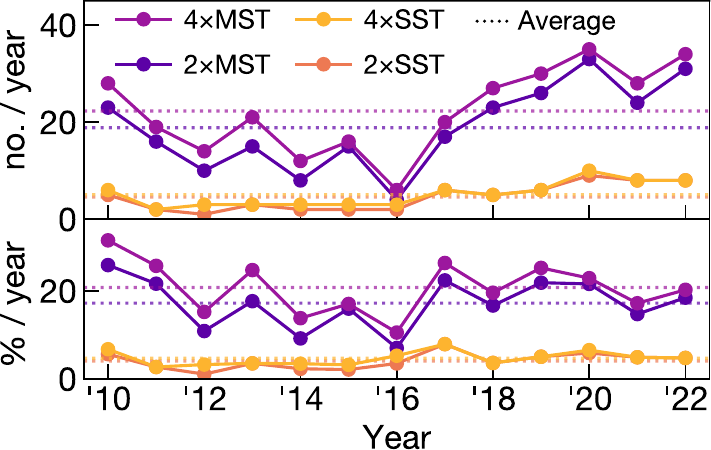}
          \caption{Number and percentage of AGN flares detectable at >$5\sigma$ that pass the $4\sigma$/TS>18 high-energy FAVA cut (alternate of \autoref{thebigone}, left).}
          \label{thebigone_b}
    \end{center}
\end{figure}

\begin{figure}[h!]
    \begin{center}
          \centering
          \includegraphics[width=1\columnwidth]{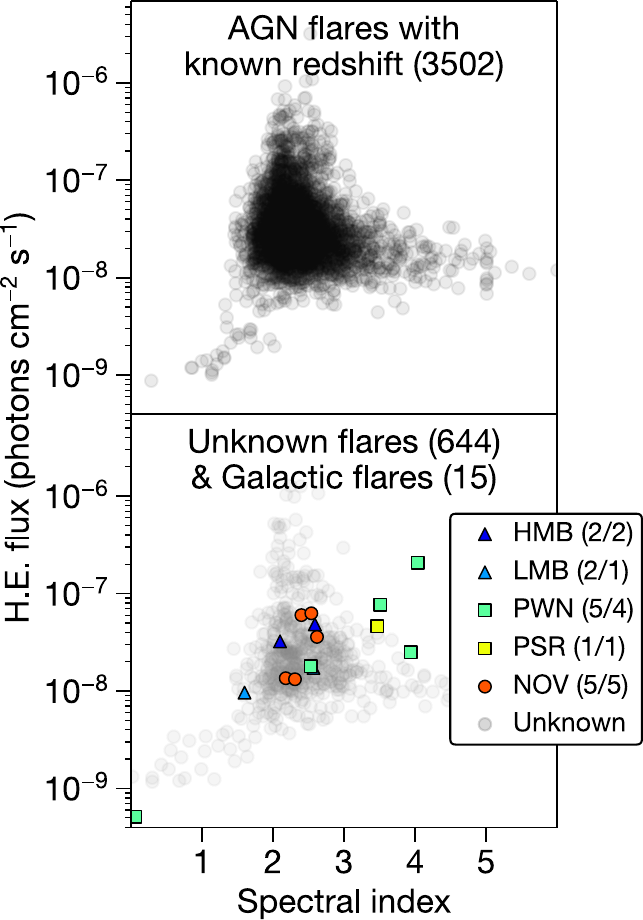}
          \caption{Photon flux and spectral index of high-energy FAVA flares that pass the $4\sigma$/TS>18 high-energy FAVA cut (alternate of \autoref{sources}).}
          \label{sources_b}
    \end{center}
\end{figure}

\section{GRB modelling}

\begin{figure}[h!]
    \begin{center}
          \centering
          \includegraphics[width=0.75\columnwidth]{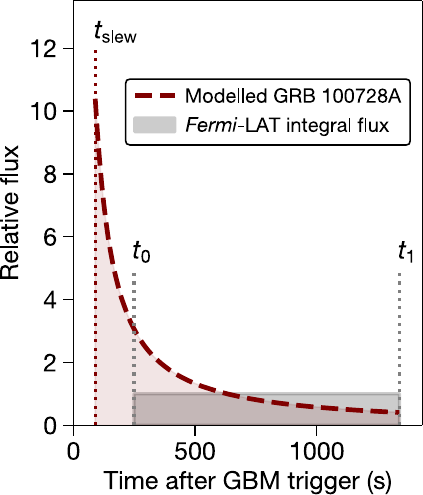}
          \caption{Demonstration of the GRB modelling method, using GRB 100728A as an example.
          $t_{\mathrm{slew}}$ is the maximum slew time of an MST, $t_0$ is the duration after the \emph{Fermi}-GBM trigger at which \emph{Fermi}-LAT started observing, and $t_1$ is when it finished.
          }
          \label{grb_cartoon}
    \end{center}
\end{figure}

The GRBs were modelled with a temporal decay of $t^{-1.2}$.
Observations were simulated to start at the maximum slew time of the telescope ($t_{\mathrm{slew}}$ seconds after the trigger from \emph{Fermi}-GBM).
\autoref{grb_cartoon} shows that the flux was scaled such that the integral flux of the decaying GRB from the start time of the \emph{Fermi}-LAT observation ($t_0$) to the end ($t_1$) was equal to the integral flux as measured by \emph{Fermi}-LAT.

\section{CTA telescope details} \label{cta_appendix}

The Small-Sized Telescope (SST) is a dual-mirror telescope design with a 4.3\,m diameter primary reflector and 1.8\,m secondary reflector \citep{ctao2023}.
Its optimal sensitivity range is from 5--300\,TeV and it has a field-of-view of 8.8\textdegree.
The CHEC SST camera simulated in this study is a 2048 pixel camera with silicon photomultiplier sensors, sampling images at a rate of 1\,GHz.
\newline
\newline
The Medium-Sized Telescope (MST) is a single-mirror telescope design with an 11.5\,m diameter reflector \citep{ctao2023}.
Its optimal sensitivity range is from 150\,GeV--5\,TeV, and it has a field-of-view of 7.5\textdegree\ or 7.7\textdegree\ when using the FlashCam or NectarCAM camera, respectively.
The FlashCam camera simulated in this study is being used for the MSTs at the CTA-South site.
The camera has a hexagonal grid of 1764 photomultiplier pixels, sampling images at a rate of 250\,MHz through a purely digital pipeline.
\end{appendix}

\end{document}